\documentclass{emulateapj}
\usepackage{graphicx}
\usepackage{epstopdf}
\begin{document}
\submitted{ApJ Accepted December 18 2009}
\shorttitle{A Nearby Blue L Dwarf From SDSS}
\title{SDSS J141624.08+134826.7: A Nearby Blue L Dwarf From the \\Sloan Digital Sky Survey}
\author{Brendan P. Bowler, Michael C. Liu, and Trent J. Dupuy}
\affil{Institute for Astronomy, University of Hawai`i \\ 2680 Woodlawn Drive, Honolulu, HI 96822, USA}
\email{bpbowler@ifa.hawaii.edu}

\altaffiltext{1}{Visiting Astronomer at the Infrared Telescope Facility, which is operated by the University of Hawaii under Cooperative Agreement no. NCC 5-538 with the National Aeronautics and Space Administration, Science Mission Directorate, Planetary Astronomy Program.}

\begin{abstract}

We present the discovery of a bright ($J$ = 13.1 mag) nearby L6 dwarf found in a search for L-type ultracool subdwarfs in the Sloan Digital Sky Survey (SDSS) Data Release~7.   SDSS J141624.08+134826.7 exhibits blue near-infrared colors compared to other optically-typed L6 objects, but its optical and near-infrared spectra do not show metal-poor features characteristic of known L-type ultracool subdwarfs.  Instead, SDSS J141624.08+134826.7 is probably a nearby example of the class of L dwarfs with low condensate opacities which exhibit unusually blue near-infrared colors for a given spectral type.  Its deep 1.4 and 1.9~$\mu$m H$_2$O  absorption bands, weak 2.3~$\mu$m CO feature, strong 0.99~$\mu$m FeH band, and shallow optical TiO and CaH bands resemble the spectra of other blue L dwarfs which are believed to have unusually thin or large-grained cloud structure.  The luminosity of SDSS J141624.08+134826.7 implies that it is either a high-mass brown dwarf or a low mass star, depending on its age, and its $UVW$ space motion suggests a thin-disk membership.  With a spectrophotometric distance of 8.4~$\pm$~1.9 pc, SDSS J141624.08+134826.7 is one of the nearest L dwarfs to the Sun and is therefore an excellent target for high resolution imaging, spectroscopic, and astrometric follow-up observations.

\end{abstract}
\keywords{stars: low-mass, brown dwarfs ---   stars: individual (SDSS J141624.08+134826.7)}

\section{Introduction}



Over the past 15 years several hundred L and T dwarfs have been found within $\sim$~100~pc of the Sun.  Untangling the atmospheric and evolutionary properties of these objects as a function of mass, age, and metallicity is a leading goal of brown dwarf astrophysics.  The rarest classes of ultracool dwarfs are particularly insightful because their extreme properties map the diversity of physical parameters that exist in nature.

One such class are the blue L dwarfs that have been identified by their outlying $J$ -- $K_\mathrm{S}$ colors compared to normal L dwarfs (e.g., \citealt{Cruz:2003p67}; \citealt{Knapp:2004p15209};  \citealt{Folkes:2007p19476}; \citealt{Burgasser:2008p3725}).  Several explanations have been invoked to account for these blue NIR colors.  A reduced metallicity manifests as supressed NIR flux longward of $\sim$ 1.2 $\mu$m as a result of increased collision-induced absorption by H$_2$ (\citealt{Linsky:1969p3947}; \citealt{Borysow:1997p3945}), creating blue NIR colors.  Unresolved cool companions may also produce abnormally blue NIR colors by adding disproportionally more $J$-band flux to a composite spectrum, especially when the primary is a late-M or L dwarf and the unseen companion is a T dwarf (e.g., \citealt{Cruz:2004p17677}; \citealt{Liu:2006p14530}; \citealt{Burgasser:2008p14471}).    Finally, variations in cloud structure are thought to affect the $J$~--~$K_\mathrm{S}$ colors of L dwarfs by altering the condensate opacity (\citealt{Knapp:2004p15209}; \citealt{Burgasser:2008p3725}). 

In this Letter we report on the serendipitous discovery of SDSS J141624.08+134826.7 (hereafter SDSS J1416+1348), a bright L dwarf with blue NIR colors found in a search for ultracool subdwarfs in the Sloan Digital Sky Survey (SDSS; \citealt{York:2000p19462}).  Its spectrophotometric distance places it within 10 pc of the Sun.

\section{A Search for L-Type Subdwarfs with SDSS}

SDSS is an ongoing optical photometric and spectroscopic survey using a dedicated 2.5-m telescope at Apache Point Observatory.  The seventh data release (DR7; \citealt{Abazajian:2009p18572}) recently marked the end of the second phase of the survey, SDSS II, and includes $\sim$ 11,600 deg$^2$ of imaging data in $u$, $g$, $r$, $i$, and $z$ photometric bands.  Medium resolution (R $\equiv$ $\lambda$/$\Delta \lambda$ $\sim$ 2000) spectroscopic observations from 3800-9200 \AA \ were obtained for $\sim$ 9,400 deg$^2$ of sky for $\sim$ 460,000 stars.

The SDSS photometric and spectroscopic database is a rich resource for identifying new cool and ultracool dwarfs (e.g., \citealt{Leggett:2000p19460}, \citealt{Chiu:2006p17707}, \citealt{Bochanski:2007p19461}).  Recently \citet{Sivarani:2009p15645} found a peculiar L-type object in the SDSS spectral database, SDSS J125637.13-022452.4 (SDSS J1256-0224, sdL3.5), whose unique spectral, photometric, and kinematic properties suggest the object is an inner halo ultracool subdwarf (see also \citealt{Burgasser:2009p18452}).  Metal-poor L dwarfs are quite rare in the solar neighborhood; only three other sdL-type objects are currently known (\citealt{Burgasser:2003p577}, \citealt{Burgasser:2004p564}, \citealt{Cushing:2009p17782}).  

In an effort to identify more L-type ultracool subdwarfs, we searched the SDSS spectral database for objects with similar colors and spectral features to SDSS J1256-0224.  We selected all objects with late-type stellar spectral flags (SpecClass = ``Star\_Late'') within the following color and magnitude limits (AB mags): 1.5 $<$ $i$ -- $z$ $<$ 3.5,  0.0 $<$ $r$ -- $i$ $<$ 6.0, and $i$ $<$ 20.0.  The color selection criteria were defined to include objects that are slightly bluer than SDSS J1256-0224 ($i$--$z$ = 1.7, $r$ -- $i$ = 2.4) and to allow for much redder red-optical colors.  The magnitude criterion was set to exclude noisy spectra.  

Altogether 1581 objects satisfied these criteria.  We visually inspected each spectrum in search of L-type objects with strong TiO and CaH absorption features at $\sim$ 7100 \AA \ and 6800 \AA, respectively, which are characteristic of known L-type subdwarfs.  These deep features contrast sharply with the shallow CaH and TiO absorption depths shortward of 7500 \AA \ in the optical spectra of normal L dwarfs.  Most of the objects in our sample were late-type M dwarfs with a small fraction being L dwarfs.  We recovered the sdL3.5 object SDSS J1256-0224 but failed to find other similar ultracool subdwarfs.  

We did, however, uncover a previously unknown bright late-type L dwarf, SDSS J1416+1348, which is the subject of this Letter.  Its peculiar blue NIR colors are coincidentally characteristic of the ultracool subdwarfs we were searching for, but they were not a part of our search criteria.  The SDSS fiber-fed spectrum of SDSS J1256-0224 was obtained on 13 March 2005 UT and has a median signal-to-noise ratio of 14 between 6300-9200 \AA.

We obtained a low resolution ($R$ $\sim$ 100) 0.75-2.5 $\mu$m spectrum of SDSS J1416+1348 using the prism mode of the SpeX spectrograph (\citealt{Rayner:2003p2588}) mounted on the 3-m NASA Infrared Telescope Facility (IRTF) on the night of 4 April 2009 UT.  The weather was poor with variable cirrus clouds and $\sim$ 1$\farcs$3 FWHM seeing. We obtained a total of 1440 s of exposure time with the 0$\farcs$8 slit width at an airmass of 1.03.  Immediately afterwards we observed the A0V star HD 131951 for telluric correction.  The data were reduced using version 3.4 of the Spextool data reduction package (\citealt{Vacca:2003p497}, \citealt{Cushing:2004p501}).  The median signal-to-noise ratio of our final spectrum is 45 but reaches in excess of 100 in the $J$ and $H$ bandpasses.

\section{Analysis}{\label{section:analysis}

\subsection{Spectral Classification}

The optical spectrum of SDSS J1416+1348 is presented in Figure \ref{f1.eps} along with the spectra of L5-L8 dwarf templates from \citet{Kirkpatrick:1999p17664}.  The optical spectrum of SDSS J1416+1348 is most similar to the L6 dwarf 2MASS J0850359+105716 (2MASS J0850+1057) from 6300-9200~\AA, although a good match to the L7 dwarf DENIS-P J0205.4--1159 is achieved from 6300-8000~\AA.  Based on the similarity to 2MASS J0850+1057, we assign SDSS J1256-0224 an optical spectral type of  L6.0 $\pm$ 0.5.

Like field L dwarfs, SDSS J1416+1348 exhibits weak absorption features redward of 7500~\AA.  This is in contrast to the optical spectra of known L subdwarfs which exhibit deep TiO and CaH absorption bands from 6700-7200~\AA.  These enhanced features that are thought to be caused by a reduced metallicity (\citealt{Bessell:1982p13345}).  The optical TiO and CaH bands SDSS J1416+1348 appear to be slightly deeper than the field L dwarfs in Figure \ref{f1.eps} and may be a sign that SDSS J1416+1348 is slightly metal-poor.  

No \ion{Li}{1} is observed in SDSS J1416+1348 (EW $\lesssim$ 2 \AA) which suggests that it is a low-mass star or an old high mass brown dwarf by the lithium test (\citealt{Magazzu:1993p19478}), although another possibility is that  SDSS J1416+1348 is a brown dwarf with a mass below the Li burning limit ($\sim$ 0.06 M$_{\odot}$; \citealt{Burrows:2001p64}) but with a temperature cool enough ($\sim$ 1300-1600 K; \citealt{Lodders:1999p19479}; \citealt{Burrows:1999p19480}) for lithium-bearing molecules like LiCl to form.  \citet{Weck:2004p19483} derive mid-infrared LiCl opacities for brown dwarfs and show that LiCl may be detectable in late-L dwarfs at the level of a few percent at 15.8~$\mu$m.  The brightness of SDSS J1416+1348 ($\S$\ref{section:distance}) presents an excellent opportunity to search for LiCl; its detection would indicate that SDSS J1416+1348 is a brown dwarf while a non-detection would suggest that it is a high-mass brown dwarf or low-mass star.  

Figure \ref{f2.eps} shows our SpeX NIR spectrum of SDSS J1416+1348.  As with normal L dwarfs, SDSS J1416+1348 exhibits deep H$_2$O absorption bands at 1.4 and 1.9 $\mu$m, CO absorption at 2.3 $\mu$m, and various atomic and molecular features blueward of 1.3 $\mu$m.  Overplotted in the same figure is the NIR spectrum of the L6 dwarf 2MASS J10101480--0406499 (2MASS J1010--0406) from  \citet{Reid:2006p19467}, whose $J$ -- $K_\mathrm{S}$ color of 1.9 mags places it near the mean $J$ -- $K_\mathrm{S}$ color of L6 dwarfs (Figure \ref{f3.eps}, top panel).  The most obvious difference between SDSS J1416+1348  and 2MASS J1010--0406 is the reduced flux of SDSS J1416+1348  longward of $\sim$ 1.3 $\mu$m.  Other notable differences include significantly stronger 1.4 and 1.9 $\mu$m H$_2$O absorption bands in SDSS J1416+1348, a weakened CO bandhead at 2.3 $\mu$m, a strengthened FeH feature at 0.99 $\mu$m, and the absence of the 1.6 $\mu$m FeH absorption feature which is present in 2MASS J1010--0406 and most, but not all, L dwarfs.  Also overplotted in Figure \ref{f2.eps} is the NIR spectrum of 2MASS J11263991--5003550 (2MASS J1126--5003) from \citet{Burgasser:2008p3725} (see also \citealt{Folkes:2007p19476}), which is a blue L4.5 dwarf that is believed to have thinner or larger-grained clouds compared to normal L dwarfs.  SDSS J1416+1348 has a remarkably similar spectrum to 2MASS J1126--5003 and, when normalized to 1.2-1.3 $\mu$m as in Figure \ref{f2.eps}, only diverges from 2MASS J1126--5003 at $\gtrsim$ 2 $\mu$m.

We use the indices of \citet{Reid:2001p4858} and \citet{Geballe:2002p19463} to derive a NIR spectral type for SDSS J1416+1348.  The K1, H$_2$O$^A$, and H$_2$O$^B$ indices from \citet{Reid:2001p4858} yield NIR spectral types of L3.0 $\pm$ 1.0, L5.5 $\pm$ 3.5, and L5.5 $\pm$ 2, respectively.  The 1.5 $\mu$m H$_2$O and 2.2 $\mu$m CH$_4$ indices from \citet{Geballe:2002p19463} yield NIR spectral types of L7.0 $\pm$ 1.0 and L7.5 $\pm$ 1.0.  We compute uncertainties for the indices in a Monte Carlo fashion, incorporating spectral measurement errors and the intrinsic scatter in the $SpT$-index relations.  The wide range of classifications reflects the peculiar spectrum of SDSS J1416+1348.  We adopt the average and rms of the above values, L6pec $\pm$ 2, as the NIR spectral type.

\subsection{Photometric Distance}{\label{section:distance}

We estimate a distance of 7.0 $\pm$ 1.4 pc to SDSS J1416+1348 using the $J$-band  absolute magnitude-optical spectral type relations from \citet{Tinney:2003p14852}.  The error incorporates the uncertainty in the measured apparent magnitude, the rms uncertainty of the $M_J$-$SpT$ relation, and the uncertainty in the spectral type.  The $K_\mathrm{S}$-band spectrophotometric distance of 9.7 $\pm$ 1.9 pc is consistent with the value derived using the $J$-band magnitude.  However, because the absolute magnitude-$SpT$ relations were derived using ``normal'' L dwarfs, these relations are probably not as accurate for blue L dwarfs.   We adopt the mean and rms of the $J$- and $K_\mathrm{S}$-band spectrophotometric distances, 8.4 $\pm$ 1.9 pc, as our distance estimate for SDSS J1416+1348.  

If the blue NIR colors of SDSS J1416+1348 are caused by reduced metallicity then we should employ subsolar metallicity absolute magnitude-$SpT$ relations to estimate its distance.  Using the relations from \citet{Cushing:2009p17782} we derive $J$-, $H$-, and $K_\mathrm{S}$-band spectrophotometric distances of 10.6 $\pm$ 1.0 pc, 9.5 $\pm$ 1.0, and 8.1 $\pm$ 1.0 pc, respectively, with an unweighted mean and rms of 9.4 $\pm$ 1.3 pc, which is consistent with our solar-metallicity estimates.  Even if SDSS J1416+1348 is a metal-poor L dwarf then it still appears to be located within 10 pc from the Sun (see also $\S$ 3.4).

\subsection{Physical Properties}

We fit the AMES-Dusty atmospheric models (\citealt{Allard:2001p14776}) to our IRTF/SpeX NIR spectrum to estimate the physical parameters of SDSS J1416+1348.\footnote{We obtained poor fits using the PHOENIX-$GAIA$ atmospheric models (\citealt{Brott:2005p301}) which do not include the effects of dust.}  The AMES-Dusty models consider the limiting case of dust in chemical equilibrium with the gas phase.  We fit the solar-metallicity grid of synthetic spectra spanning 1500 K $\le$ $T_\mathrm{eff}$ $\le$ 3400 K ($\Delta$$T_\mathrm{eff}$ = 100 K) and 4.0 $\le$ log $g$  (cgs) $\le$ 6.0 ($\Delta$log $g$ = 0.5) using the method described in \citet{Bowler:2009p19445}.  The best-fitting model has an effective temperature of 2200 K and a surface gravity of 5.5 (hereafter written as [2200/5.5]; Figure \ref{f3.eps}, inset).  For comparison we fit the NIR spectrum of 2MASS J1010--0406 (displayed in Figure \ref{f2.eps}), which has the same optical spectral type as SDSS J1416+1348 (L6) but has considerably redder NIR colors and is representative of ``normal'' L6 dwarfs.  The best fitting model to 2MASS J1010--0406 is [1800/5.5]; the blue NIR spectrum of SDSS J1416+1348 yields a best-fitting model which is 400 K warmer than for normal L dwarfs.  We obtain the same best-fitting models (as cited above) when we exclude the 1.5-1.7 $\mu$m spectral region, which contains an FeH feature that is missing from the model line lists.

We combine our IRTF/SpeX near-IR spectrum with the best-fitting
[2200/5.5] AMES-Dusty model to construct the complete spectral energy
distribution (SED).  We flux-calibrate the SED using the 2MASS
$K_\mathrm{S}$-band photometry and our photometric distance.  We then
integrate the SED to measure the bolometric luminosity ($L_{\rm bol}$) using
a Monte Carlo approach to account for the measurement errors (from the
finite S/N of the NIR spectrum, the 2MASS-based flux calibration, and
the photometric distance).  To gauge the systematic impact of a
mismatched atmospheric model we also use the [1800/5.5] AMES-Dusty
model and find a $\sim$0.02~dex resulting uncertainty in
$\log(L_\mathrm{bol}/L_{\odot}$).  Including the error in the
distance, we find $\log (L_\mathrm{bol}$/$L_{\odot}$) = --4.36~$\pm$~0.21~dex.\footnote{Our $L_\mathrm{bol}$ measurement from the entire SED agrees
 well with that inferred from using the $K$-band bolometric
 corrections ($BC_K$) from \citet{Golimowski:2004p15703} as revised by
 Liu, Leggett \& Dupuy (2009, submitted).  Using $BC_K$ with the
 observed $K$-band magnitude and L6$\pm$0.5 spectral type gives
 $\log(L_\mathrm{bol}/L_{\odot}$) = --4.42 $\pm$ 0.04 (0.21) dex, if one ignores
 (includes) the distance error.}

Based on the solar-metallicity evolutionary models of \citet{Burrows:1997p2706}, 
we use the measured $L_\mathrm{bol}$ and assumed ages of 1, 3, and
10~Gyr to determine masses for SDSS~J1416+1348 of $61\pm9$, $78\pm3$,
and $80.9\pm1.2$~$M_\mathrm{Jup}$.  Thus, the object appears to be right at the
stellar/substellar boundary and is either a high-mass brown dwarf or a
low-mass star assuming it is a single object.  These masses are consistent with the non-detection of
lithium in the optical spectrum.

\subsection{Space Motion}

The space motion of SDSS J1416+1348 can provide a clue about the physical origin of its blue NIR colors.  \citet{Faherty:2008p14766} analyzed the kinematics of over 800 ultracool dwarfs and found that L dwarfs with unusually blue colors have kinematics consistent with the Galactic thick disk or halo.  This suggests that metallicity and/or high surface gravity is the culprit of the blue colors for that population.  Similarly, most ultracool subdwarfs discovered to date have thick disk or halo orbits (e.g., \citealt{Burgasser:2003p577}; \citealt{Cushing:2009p17782}).



We measure the proper motion of SDSS J1416+1348 using relative astrometry
derived from the imaging surveys DSS-2 ($RI$), 2MASS ($JHK$), SDSS
($riz$), and UKIDSS\footnote{Details about the UKIDSS project can be found in \citet{Lawrence:2007p19469}, \citet{Casali:2007p19471}, \citet{Hewett:2006p19472}, \citet{Hodgkin:2009p17656}, and \citet{Hambly:2008p19468}. We have used data from the 6th Data Release.} ($YJHK$).  We use SExtractor (\citealt{Bertin:1996p19827}) 
to measure the positions of all objects in the original
survey images and SCAMP (\citealt{Bertin:2006p19828}) to determine an astrometric
solution for each data set.  After matching common objects between
data sets, we compute the positional offset of SDSS J1416+1348 relative to
well-detected (S/N $>$ 10) reference stars within 3$\arcmin$.  There
are 4 to 21 such reference stars for every pair of
images. For the relative astrometry between any two epochs, we
adopt the mean and standard deviation of the offsets computed for all common reference
objects.  We validate these astrometric errors, which range from $\sim$~30~mas
for SDSS and UKIDSS to $\sim$ 300~mas for DSS-2, by confirming that
our resulting fit to the motion of SDSS J1416+1348 has a reduced $\chi^2$ value near
unity.  We find that we also need to include parallax motion in
order to achieve a reasonable reduced $\chi^2$, as the high quality UKIDSS and
SDSS data show a clear deviation from simple proper motion.  This is
not surprising given SDSS J1416+1348's spectrophotometric distance of $\sim$ 8.4~pc.  We fit for
the proper motion and parallax using all $N(N-1)/2$ possible epoch
pairs, employing the Levenbarg-Marquardt algorithm for least-squares
minimization to perform the fit. To assess the uncertainty in our
best-fit parameters, we apply the same fitting procedure to $10^3$
simulated astrometric data sets at the same epochs but with randomly
drawn noise added.  We measure a proper motion of $\mu$ = 151 $\pm$
8~mas/yr at a position angle of 33 $\pm$ 4$^{\circ}$.  This proper motion is consistent with fitting for 
the proper motion only based on catalog positions with larger uncertainties.  Although including
parallax was necessary to accurately model the data, our simulations
show that it is not measured at a high precision ($\pi_{\rm rel} = 107
\pm 34$~mas).  Thus, we continue to adopt the more precise photometric
distance estimate of 8.4 $\pm$ 1.9~pc, which is consistent with our astrometric
analysis.

Our proper motion measurement and distance estimate can be combined with the radial velocity of SDSS J1416+1348 to derive $UVW$ kinematics. The radial velocity reported by SDSS is  --88 $\pm$ 33 km s$^{-1}$, which is derived by cross correlation with a late-type stellar spectrum.  To improve upon this, we fit Gaussians to the \ion{Rb}{1} (7800, 7948 \AA), \ion{Na}{1} (8183, 8195 \AA), and \ion{Cs}{1} (8521, 8943 \AA) absorption lines, convert line centers from heliocentric vacuum wavelengths to air wavelengths using the IAU standard conversion in \citet{Morton:1991p19487}, and compute a radial velocity using the rest wavelengths in \citet{Ralchenko:2008p19486}.  We obtain internally consistent values with a mean and rms velocity of --38 $\pm$ 10 km s$^{-1}$, which is consistent with the SDSS radial velocity at the 2 $\sigma$ level.  Using our solar-metallicity spectrophotometric distance estimate of 8.4 $\pm$ 1.9 pc and our radial velocity measurement we find ($U$, $V$, $W$)$_\mathrm{LSR}$ velocity components of (6 $\pm$ 4 km s$^{-1}$, 10.2 $\pm$ 1.4 km s$^{-1}$, --27 $\pm$ 9 km s$^{-1}$).  Here $U$ is positive towards the galactic anticenter.  The velocities are calculated following \citet{Johnson:1987p19474} and are corrected for the solar motion with respect to the local standard of rest using the values from \citet[$UVW_{\odot}$ = \{--10.00, +5.25, +7.17\}]{Dehnen:1998p19473}. Errors are computed in a Monte Carlo fashion from 10$^6$ trials and are dominated by the radial velocity uncertainty.  

We derive the probability that SDSS J1416+1348 is a member of the thin disk, thick disk, or halo kinematic population using the Besan\c{c}on Galactic model (\citealt{Robin:2003p19488}).  The $UVW$ velocities of SDSS J1416+1348 imply a 99.1\% (0.9\%) probability of having a thin (thick) disk membership, with a negligible halo membership, based on the method discussed in  \citet{Dupuy:2009p18533}.   Similarly, the Toomre diagram from \citet[note the galactic standard of rest]{Venn:2004p19475} suggests a Galactic thin disk orbit.

\section{Discussion}{\label{section:discussion}}

Three mechanisms have been proposed to cause the peculiar NIR colors of blue L dwarfs: low metallicity, unresolved multiplicity, or thin/large grain condensate clouds.  As discussed in \citet{Burgasser:2008p3725}, a high surface gravity will also cause blue NIR colors in L dwarfs, but current model atmospheres with high surface gravity fail to reproduce the deep H$_2$O absorption features observed in many blue L dwarfs.  A significantly reduced metallicity for SDSS J1416+1348 is unlikely based on the shallow depth of the metallicity-dependent TiO and CaH absorption bands in its optical spectrum.  Similarly, the NIR spectrum of SDSS J1416+1348 does not exhibit the extreme collision-induced H$_2$ absorption seen in the spectra of known L subdwarfs.  All known L subdwarfs also have inner or outer halo orbits, but the kinematics of SDSS J1416+1348 implies a thin disk orbit.  Another possibility is that SDSS J1416+1348 is only mildly metal-poor, but slightly low-metallicity atmospheric models do not predict the deep NIR H$_2$O bands observed in SDSS J1416+1348 (\citealt{Burgasser:2008p3725}; \citealt{Burrows:2006p7009}).

\citet{Burgasser:2008p3725} found that unresolved multiplicity is not likely to be the cause of the blue NIR colors of 2MASS J1126--5003 based on composite binary spectral template fitting.  Because of the similarity of the NIR spectra of SDSS J1416+1348 and 2MASS J1126--5003 (Figure \ref{f2.eps}), and because the spectrum of SDSS J1416+1348 does not show the characteristic feature of unresolved M/L + T binaries at 1.6 $\mu$m (\citealt{Burgasser:2008p14471}), it is also unlikely that unresolved multiplicity is the cause of the blue NIR colors of SDSS J1416+1348.

We find that the most likely explanation for the peculiar colors of SDSS J1416+1348 is a reduced condensate opacity caused by thin or large-grain clouds.  This is supported by the abnormally strong 1.4 and 1.9 $\mu$m H$_2$O features, which can only be reconciled with atmospheric models that contain large grains (\citealt{Burrows:2006p7009}) and/or thin clouds (\citealt{Stephens:2009p19484}).  The 0.99 $\mu$m FeH feature of SDSS J1416+1348 has almost exactly the same depth as in 2MASS J1126--5003, both of which are significantly stronger than in normal L dwarfs.  As discussed in \citet{Folkes:2007p19476} and \citet{Burgasser:2008p14471}, this deep FeH is further evidence for a peculiar cloud structure in blue L dwarfs.

It appears that previous searches for L dwarfs missed  SDSS J1416+1348 because of its small proper motion and  peculiar colors, despite its bright NIR magnitudes.  Recently \citet{Zhang:2009p19465} performed a search for ultracool dwarfs in SDSS DR7 but SDSS J1416+1348 was missed because of their bright $i$- and $z$-band magnitude cuts.  Many of the  SDSS-only and SDSS cross-matching searches were performed with early data releases which did not include SDSS J1416+1348 (e.g., \citealt{Chiu:2006p17707}), as SDSS J1416+1348 has imaging observations from May 2003 and spectroscopic observations from March 2005.   \citet{Cruz:2003p67} performed a series of 2MASS, Tycho, and USNO  color and magnitude cuts to search for L dwarfs within 20 pc (also see footnote 14 in \citealt{Cruz:2007p19477}).  SDSS J1416+1348 appears to pass the magnitude and color criteria in that search so it is unclear why it was not found.  SDSS has surveyed nearly one quarter of the sky; because only one bright, blue L dwarf with a small proper motion has been found to date, we suspect that at most a handful of objects like SDSS J1416+1348 await discovery.  
 
Based on its spectrophotometric distance,  SDSS J1416+1348 is one of the nearest L dwarfs to the Sun (Figure \ref{f3.eps}, bottom panel; see also \citealt{Faherty:2008p14766}) and with a $J$-band magnitude of 13.148 it is one of the brightest late-type L dwarfs known.  Its proximity makes it an excellent target for follow-up parallax measurements, radial velocity measurements, adaptive optics imaging observations, and high resolution, high signal-to-noise optical and NIR spectroscopic observations.

 \acknowledgments

We thank Michael Cushing for his helpful comments on a draft of this article.  This research has benefitted from the SpeX Prism Spectral Libraries, maintained by Adam Burgasser at http://www.browndwarfs.org/spexprism.  We have also made use of the SIMBAD database, the Two Micron All Sky Survey, SDSS, and the UKIRT Infrared Deep Sky Survey.  

\newpage

 \bibliographystyle{apj}

\begin{figure}
  \resizebox{\textwidth}{!}{\includegraphics{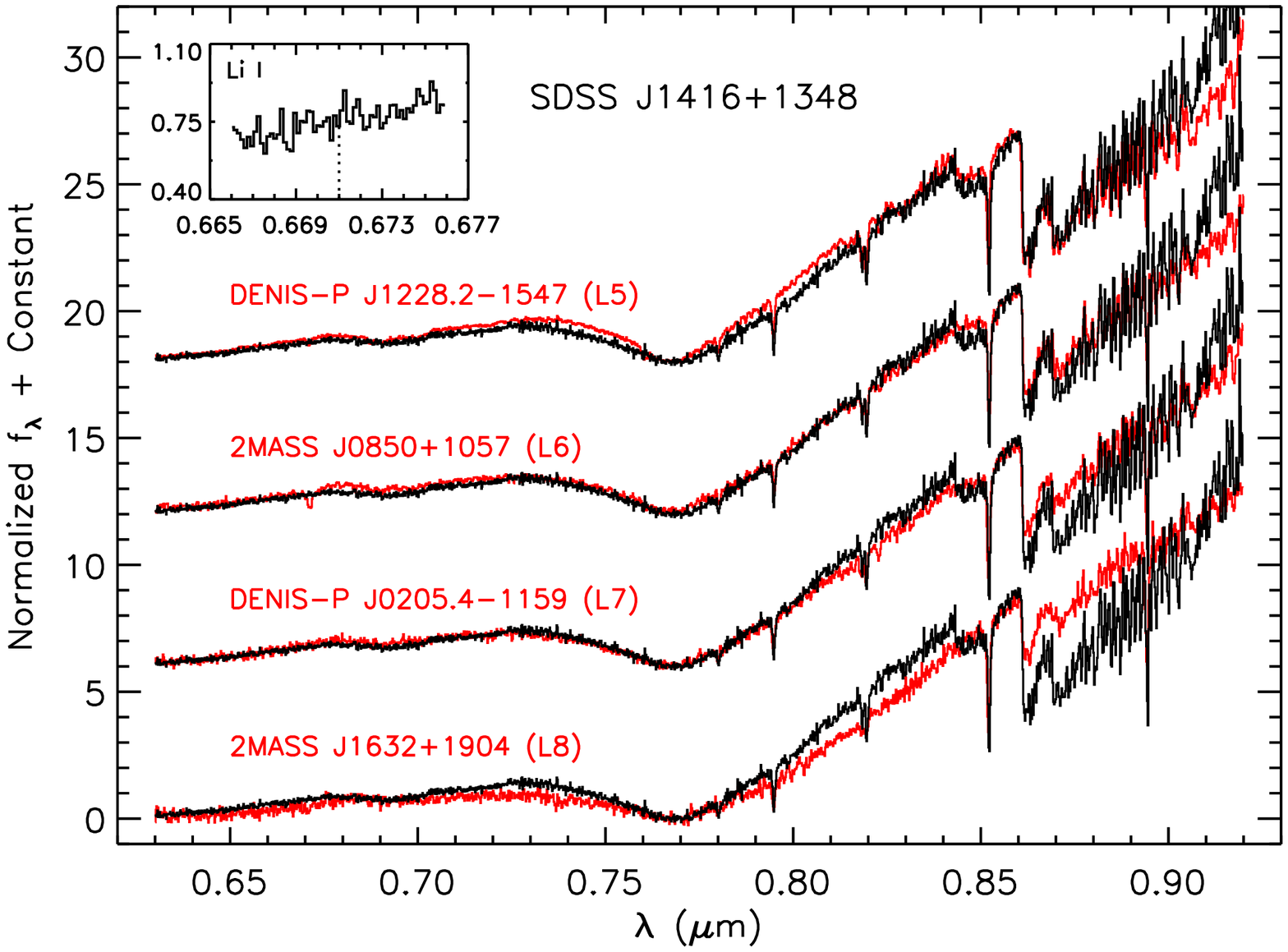}}
  \caption{SDSS optical spectrum of SDSS J1416+1348 (black).  Overplotted are the spectra of L5-L8 dwarf templates from  \citet[red]{Kirkpatrick:1999p17664}.  SDSS J1416+1348 is most similar to the L6 dwarf 2MASS J0850+1057.  The inset shows the nondetection of \ion{Li}{1} at 6708 \AA. All spectra are normalized to unit area between 6300-9200 \AA \ and are offset for clarity. \label{f1.eps} } 
\end{figure}

\begin{figure}
  \resizebox{\textwidth}{!}{\includegraphics{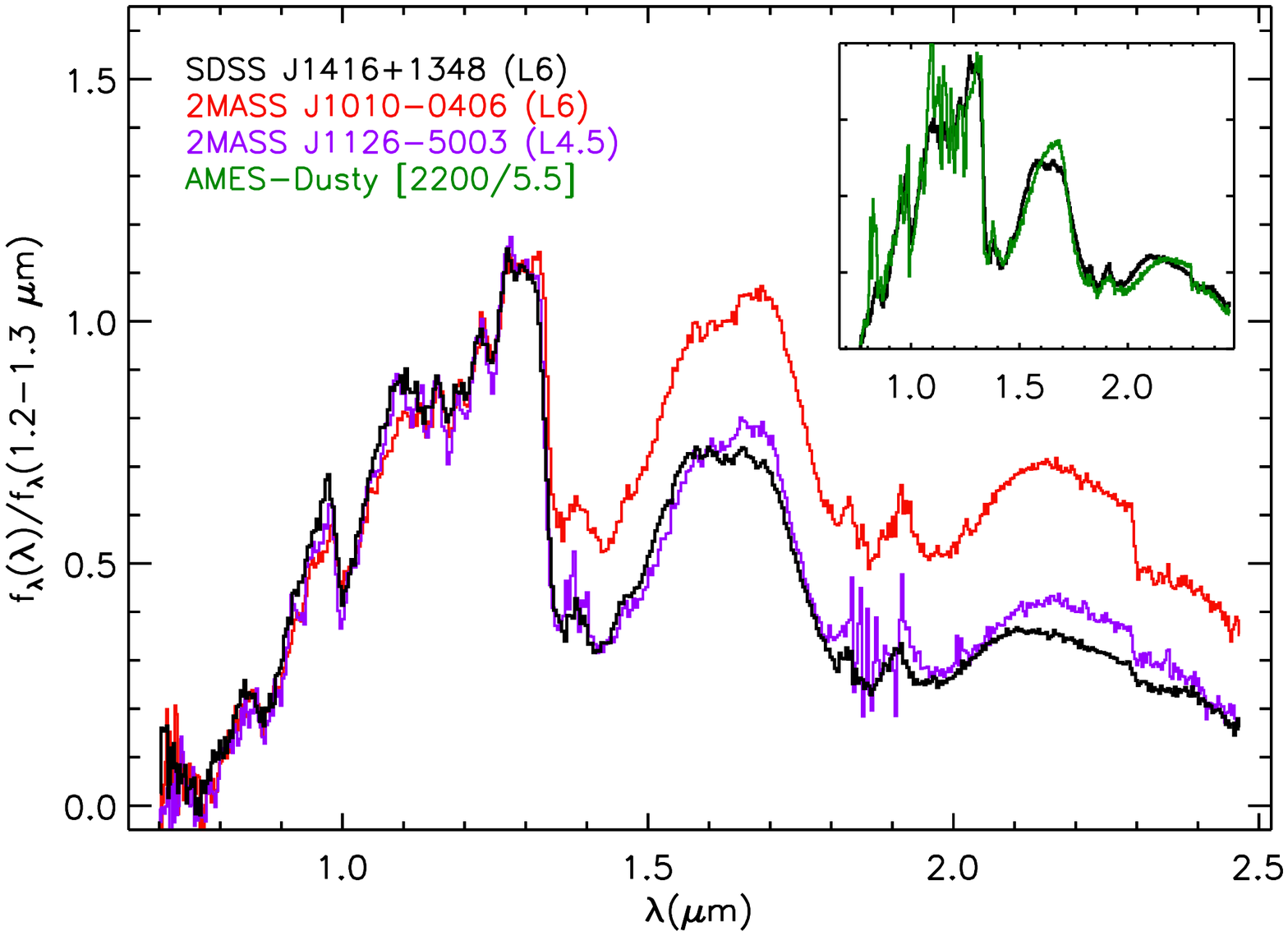}}
  \caption{Near infrared spectrum of SDSS J1416+1348 from IRTF/SpeX in prism mode (black).  Overplotted are the SpeX/prism spectra of the L6 dwarf 2MASS J1010--0406 (\citealt{Reid:2006p19467}, red), which has $J$-$K_\mathrm{S}$ colors similar to normal L dwarfs ($\sim$ 1.9 mags), and the blue L4.5 dwarf 2MASS J1126--5003 (\citealt{Burgasser:2008p3725}, purple).   All spectra are normalized to the same flux from 1.2-1.3 $\mu$m.  The inset shows the best-fitting AMES-Dusty model with $T_\mathrm{eff}$ = 2200 K and log $g$ = 5.5 (green). \label{f2.eps} } 
\end{figure}

\begin{figure}
  \resizebox{\textwidth}{!}{\includegraphics{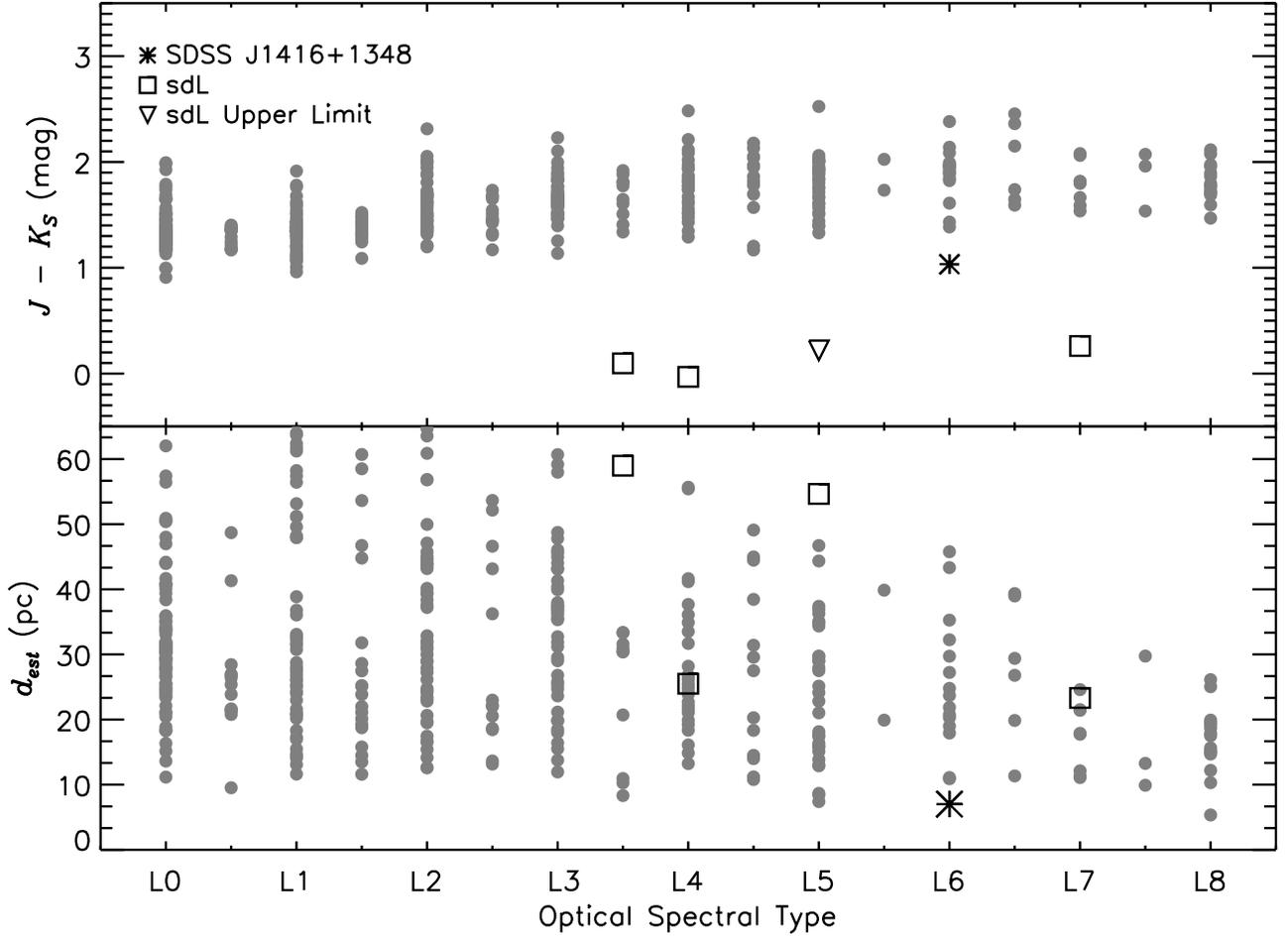}}
  \caption{\emph{Top panel:} $J$--$K_\mathrm{S}$ colors for L dwarfs from http://dwarfarchives.org as of April 2009.  L~subdwarfs are shown as open symbols while SDSS J1416+1348 is shown as an asterisk.  For late-L spectral types, blue L dwarfs have $J$--$K_\mathrm{S}$ colors of $\sim$ 1.0, while known L subdwarfs have $J$--$K_\mathrm{S}$ colors of $\sim$ 0.0.   The open squares show the colors of known L subdwarfs and the open triangle shows the upper limit for the L subdwarf 2MASS J06164006--6407194.  \emph{Bottom panel:} $J$-band spectrophotometric distances to L dwarfs from http://dwarfarchives.org.  For field L dwarfs we use the relations from \citet{Tinney:2003p14852}; for L subdwarfs we use the relations from  \citet{Cushing:2009p17782}.  Symbols are the same as in the upper panel.\label{f3.eps} } 
\end{figure}

\begin{deluxetable}{lc}
\tabletypesize{\scriptsize}
\tablewidth{0pt}
\tablecolumns{2}
\tablecaption{SDSS J1416+1348 Properties \label{properties}}
\tablehead{
        \colhead{Property}   &    \colhead{Value}   
}
\startdata
\cutinhead{Observed}
$\alpha_{J2000}$\tablenotemark{a} &  14$^\mathrm{h}$16$^\mathrm{m}$24$\fs$08 \\
$\delta_{J2000}$\tablenotemark{a} &  13$^{\circ}$48$'$26$\farcs$7 \\
Epoch   &  2003.41 \\
Optical SpT  & L6.0 $\pm$ 0.5 \\
NIR SpT  &  L6.0pec $\pm$ 2.0 \\
$\mu$ (mas yr$^{-1}$)  & 151 $\pm$ 8 \\
$PA$ ($^{\circ}$) & 33 $\pm$ 4 \\
$V_\mathrm{rad}$ (km s$^{-1}$) & --38 $\pm$ 10 \\
$u$ (SDSS) &  (23.55 $\pm$ 0.57)\tablenotemark{b}  \\
$g$ (SDSS)  &  (23.08 $\pm$ 0.18)\tablenotemark{b} \\
$r$ (SDSS)   &  20.69 $\pm$ 0.04  \\
$i$ (SDSS)   & 18.38 $\pm$ 0.01  \\
$z$ (SDSS)   & 15.92 $\pm$ 0.01   \\
$R$ (USNO-B1.0)  & 19.68 \\
$I$ (USNO-B1.0)   & 17.22 \\
$Y$ (UKIDSS) &  14.255 $\pm$ 0.003 \\
$J$ (2MASS)  &  13.148 $\pm$ 0.025 \\
$H$ (2MASS)  &  12.456 $\pm$ 0.028 \\
$K_{\mathrm{S}}$ (2MASS)  &  12.114 $\pm$ 0.023 \\

\cutinhead{Estimated}
$d_\mathrm{est}$\tablenotemark{c} & 8.4 $\pm$ 1.9 pc \\
$\log(L_\mathrm{bol}/L_{\odot})$   &  --4.36 $\pm$ 0.21 (0.02)\tablenotemark{d} \\
$U$\tablenotemark{e} (km s$^{-1}$) & 6 $\pm$ 4 \\
$V$  (km s$^{-1}$) & 10.2 $\pm$ 1.2 \\
$W$ (km s$^{-1}$) & --27 $\pm$ 9  \\

\enddata
\tablecomments{$u$, $g$, $r$, $i$, and $z$ are in AB magnitudes; $J$, $H$, and $K_{\mathrm{S}}$ are in the 2MASS system; and $Y$ is the UKIRT/WFCAM Y-band filter.}
\tablenotetext{a}{Measured by SDSS.}
\tablenotetext{b}{Near the faint limit of SDSS and may not represent an actual detection. The quoted value is the inverse hyperbolic sine magnitude ($asich$), which for high signal-to-noise measurements is the same as the canonical astronomical magnitude.  }
\tablenotetext{c}{Average of the $J$- and $K_\mathrm{S}$-band spectrophotmetric distances.}
\tablenotetext{d}{The error in parentheses excludes the uncertainty in the distance.} 
\tablenotetext{e}{Positive toward galactic anticenter.}

\end{deluxetable}

\end{document}